# Brief Report on QoSec, Context Aware Security and the Role of Physical Layer Security in 6G Wireless

**Arsenia Chorti**
Assoc. Professor, ETIS UMR8051, CY University, ENSEA, CNRS
November 2020

**1. INTRODUCTION**

While the security literature predominantly focuses on the core network, the enhancement of the security of the beyond fifth generation (B5G) access network becomes of critical importance. Despite the strengthening of 5G security protocols with respect to LTE, there are still open issues that have not yet been fully addressed. We provide in Section 2 a short review of two 3GPP reports on the "False Base Station Attack" and on the "Security Issues for URLLC" to showcase some of the open security challenges in 5G.

In parallel, proposals for the flexible allocation of the infrastructure resources under the umbrella of network slicing, bring about the need for machine learning (ML) unit enabled multi-domain orchestration. As we move gradually away from the standard client-server networking paradigm and enter a new era of truly E2E quality of service (QoS), service level agreements (SLAs) in the near future will be expected to include guarantees about the quality of security (QoSec) as well. Defining the right ingredients of QoSec, including how to identify the security level required and propose adaptive, dynamic and risk aware security solutions is currently investigated. Initial ideas about QoSec are presented in Section 3.

Meanwhile, in the radio community a different discussion is taking place around the 1949 paper of Warren Weaver on the premise that in sixth generation (6G) systems it will be possible to move towards "semantic communications", i.e., conveying reliably the "meaning" of the messages, rather than simply conveying reliably the data that carry the messages. Early approaches in this direction start from the premise that semantic content is included in the "context" of the communication, including aspects such as the location of the communicating nodes, the time, the identities of the nodes, the type of data exchanged, the reason the communications takes place. Incorporating context awareness in QoSec is projected to allow handle more efficiently aspects related to identifying the risk or threat level and the required security level. Related ideas are discussed in Section 4.

Finally, as novel sensing and intelligence capabilities are envisioned in 6G, security solutions from the palette of physical layer security can emerge, particularly for massive machine type communications involving large scale low-end IoT devices. This topic is discussed in Section 5.

In the longer 10-year horizon novel security concepts based on "trust models" and "risk-based, adaptive identity management and access control" are expected to come to life, enabled to a large extend by artificial intelligence (AI). Early elements of these directions are seen in standards used for digital (monetary) transactions, e.g., the IETF RFC Vectors of Trust. Furthermore, to allow for flexible QoSec in the differentiated services framework, the development and integration of security controls at *all* layers of the communications system is envisioned.

**2. SOME OPEN 5G SECURITY ISSUES AND 6G SECURITY CHALLENGES**

We first review a few open security issues in 5G and move on to identify potential security challenges in 6G.

### 2.1 False Base Station Attack

The expression "false base stations" (FBS) describes wireless devices that impersonate genuine base stations. FBSs are popularly known as International Number Subscriber Identity (IMSI) catchers. While one of their initial historical attacks was to catch subscribers' IMSIs, more advanced attacks have been reported since. The capabilities of FBSs vary depending on the generation of the mobile network (GPRS, UMTS, LTE, 5G). 5G, in particular, has already made significant improvements to combat FBSs. Nevertheless, there remains a number of critical security threats described in [1]; in some cases, no satisfactory solution has been identified to date. The attacks are carried out on the radio by means of overshadowing the signal emitted by legitimate BSs. In the downlink, some of the attacks are made possible by the use of unprotected messages. On the uplink, false UE (fUE) can also act as the malicious counterpart of FBSs and impersonate legitimate UEs to the legitimate serving BS. It is therefore vital to detect FBS and fUE and develop mitigation solutions to increase the robustness against these attacks.

### 2.2 Security Issues in URLLC

Furthermore, critical machine communications are typically used for industrial and other applications requiring low latency and very high reliability. To achieve high reliability, increasing diversity, e.g., via multiple parallel transmissions could be exploited; however, this consequently increases the opportunities for attacks. Furthermore, extremely low latency requires high speed authentication and re-authentication for initial access as well as very fast handover procedures. Overly aggressive latency targets could entail a new security architecture altogether. While solutions for fast authentication using implicit certificates or certificateless solutions can speed up authentication, many open challenges for sub-millisecond delay constrained URLLC systems remain, with respect not only to authentication, but as well for the integrity and confidentiality of both the control and data planes [2].

### 2.3 Jamming Attacks in mMIMO – RF Resilience

As cell and base stations grow smaller, it could become easier to launch jamming attacks. As an example, jamming during the beam allocation in mmWave [3] could even hinder the establishment of the radio link; in this aspect, standard security protocols that build on the premise that the communication link has already been established, cannot offer solutions when this is not the case. Intrusion detection at the wireless edge is required, combined with modulation choices to mitigate impact.

## NEW SECURITY CHALLENGES IN 6G

Emerging use cases in 5G, and notably massive IoT with thousands of connected devices per cell and ultra-low delays, raise new security challenges. In addition, there are security challenges related to computational capabilities; on one end we have quantum computing, on the other end SIM-less, low end IoT devices.

### 2.4 Quantum Computing

A further challenge comes from quantum computing, which has seen significant progress after massive investment to build prototypes with more than 50 qubits. The national institute of standardization (NIST) is evaluating novel post-quantum crypto algorithms to replace current state-of-the-art public key encryption schemes. Nevertheless, it is a common concern that quantum resistance will lead to an increase in terms of the complexity of the new crypto systems as key sizes might pose a significant problem.

*2.5 SIM-less IoT devices and long-term IoT security*

First of all, not all very low cost terminals will be able to support the usual security mechanisms, as they will not have the necessary computing power, and secondly, it is desirable that some of them do not have a SIM card for cost reduction reasons. Nevertheless, some of these terminals carry critical information that needs to be authenticated and protected. New protection mechanisms that are lightweight, but nevertheless as secure as the existing conventional mechanisms must therefore be put in place. Another factor at play is that IoT nodes will typically have a very long lifespan (>10 years as opposed to 3 years for a laptop) and can be distributed in large geographical areas; the impact of this is twofold. Firstly, security updates that might be necessary might be difficult to apply in large scale networks with geographically scattered IoT nodes. Secondly, despite recent advances in lightweight cryptography, it is difficult to guarantee that mass-produced, computationally and power constrained IoT devices will have a hardware capable of being updated with the necessary patches to resist all the threats that will arise in their lifetimes.

## 2. QUALITY OF SECURTITY (QoSec) (Current research)

QoSec captures the general direction of being able to provide different security level guarantees, in response to the security needs of different slices of the network, reflecting on the DiffServ QoS paradigm. A central aspect of QoSec is to identify how to make the security level adaptive, i.e., how to automatically identify the right combination of crypto schemes (encryption, integrity, authentication primitives) and how to incorporate flexibility regarding their choices in security protocols. A closely related topic is that of trust, with "zero trust" referring to providing hard security guarantees, possibly equivalent to the current definition of security level 5 (post-quantum). In future security protocols varying levels of trustworthiness (e.g., as defined by NIST in SP800-53 Rev. 4) are envisioned through the use of security control baselines. Note that security control baselines are developed based on a number of general assumptions, including common environmental, operational, and functional considerations, giving rise to the question of context awareness in security.

## 3. CONTEXT AWARENESS IN THE WIRELESS EDGE (3 to 5-year horizon)

The opening up of the THz spectrum will provide new "sensing" capabilities to 6G devices, such as high definition imaging and frequency spectroscopy. In combination with decimeter precision localization, as showcased recently for mmWave systems, these enhanced sensing capabilities can prove instrumental in understanding context and could naturally be incorporating in trust building and predicting reliability. In terms of security, in the near 3 to 5-year horizon, the following questions are to be addressed:

1. ***How to Measure the Threat Level from Context:*** PHY layer inputs, particularly in the form of sensing information including the location of a node, the time of communication, the ambient temperature, etc., carry important *contextual* information, directly related to semantics. We can envision AI based fusion of sensing information to obtain an enhanced evaluation of the *threat level.*
2. ***How to Use Context to Identify the Security Level Required for Particular Data Flows:*** We need to take steps towards defining new metrics describing the criticality of the particular data exchanged and furthermore, how valuable they are considered from an adversarial point of view. This can be thought of as the analogous of defining the priority level in QoS.
3. ***How to Match Security Levels to Security Schemes:*** After defining the security level with rapport to the context of communication, the next question is how to map this to an actual set of algorithms and security schemes. Two approaches emerge that can possibly be used jointly: i) a crypto based approaches in which the strength of crypto systems is, roughly speaking, related to the lengths of the keys (after the right transformations are accounted for); ii) physical layer security approaches in which the wireless channel and the hardware are used as sources of uniqueness (for authentication) and/or entropy (e.g., for key generation and agreement).

## 5. WHAT PHYSICAL LAYER SECURITY CAN DO FOR 6G (10-year horizon)

In the past years, physical layer security (PLS) [4], [5] has been studied and indicated as a possible way to emancipate networks from classic, complexity based, security approaches. PLS is based on the premise that we can move some of the security core functions down to the physical layer, exploiting both the communication radio channel and the hardware as sources of uniqueness or of entropy. It is usually this second aspect of PLS that is considered in the literature, around the concept of the secrecy capacity and of the secret key generation capacity. As a source of uniqueness, we can leverage PHY on the other hand by using RF fingerprinting and high precision localization and / or physical unclonable functions for authentication purposes. In essence, as the line of sight conditions and the channel quality changes, there is a clear interplay between the use of the CSI for high precision localization (i.e., as an authentication factor) or as the means to distil entropy for use in confidentiality and integrity schemes. This unique setting can only be exploited with enhanced monitoring of the wireless channel and of the context in general.

Overall, PLS can provide information-theoretic security guarantees with lightweight mechanisms (e.g., using Polar or LDPC encoders) as opposed to computationally expensive elliptic curve based cryptography. At the same time, it is more probable that PLS will be incorporated in hybrid PLS-crypto systems along with symmetric key block ciphers to sustain reasonable communications rates or will act as an extra security layer, complementing other approaches.

In the longer 10-year perspective, the foundational work of formally interconnecting PLS and semantic security can be envisioned by characterizing the predictability / unpredictability of the channel coefficient realizations in the three dimensions of time, frequency and space, as unpredictability is related to indistinguishability, a central concept in crypto proofs.